\documentclass[journal, onecolumn, draftcls]{IEEEtran}

\usepackage{cite}
\usepackage[cmex10]{amsmath}
\usepackage{graphicx}
\usepackage{url}
\usepackage{tikz}
\usepackage{tkz-euclide}

\usepackage[utf8]{inputenc}
\usepackage[normalem]{ulem}

\usetkzobj{all}
\usetikzlibrary{shapes}
\usetikzlibrary{calc}
\usetikzlibrary{3d}
\usetikzlibrary{timeline}
\definecolor{col1}{rgb}{0.0000,0.4470,0.7410}%
\definecolor{col2}{rgb}{0.8500,0.3250,0.0980}%
\definecolor{col3}{rgb}{0.9290,0.6940,0.1250}%
\definecolor{col4}{rgb}{0.4940,0.1840,0.5560}%
\definecolor{col5}{rgb}{0.4660,0.6740,0.1880}%
\definecolor{col6}{rgb}{0.3010,0.7450,0.9330}%
\definecolor{tcc}{rgb}{0,1,1}%

\begin{document}
\title{Trends and Challenges in Wireless Channel Modeling for an Evolving Radio Access}
\author{\IEEEauthorblockN{Paul Ferrand,~\IEEEmembership{Member,~IEEE}, 
Mustapha Amara,~\IEEEmembership{Member,~IEEE}}, Stefan Valentin,~\IEEEmembership{Senior Member,~IEEE} and Maxime Guillaud,~\IEEEmembership{Senior Member,~IEEE}
\thanks{The authors are with the Mathematical and Algorithmic Sciences Lab, Huawei Technologies France (emails: \{paul.ferrand, mustapha.amara, stefan.valentin, maxime.guillaud\}@huawei.com).}}

% \IEEEauthorblockA{\IEEEauthorrefmark{1}French Research Center}}% <-this % stops an unwanted space

\maketitle
\begin{abstract}
\cite{}
With the advent of 5G, standardization and research are currently defining the next generation of the radio access.
Considering the high constraints imposed by the future standards, disruptive technologies such as Massive MIMO and mmWave are being proposed.
At the heart of this process are wireless channel models that now need to cover a massive increase in design parameters, a large variety of frequency bands, and heterogeneous deployments.
This tutorial describes how channel models address this new level of complexity and which tools the community prepares to efficiently but accurately capture the upcoming changes in radio access design.
We analyze the main drivers behind these new modeling tools, the challenges they pose, and survey the current approaches to overcome them.
\end{abstract}

\section{Introduction}

The 5\textsuperscript{th} generation (5G) of mobile wireless standards will be drafted from 2020. Recently, industrial and academic teams are focusing on one of its prerequisites: agreeing on common evaluation scenarios and methodologies.
This step is crucial as it will be the foundation to evaluate and compare all proposals to the 5G standards.
One key aspect in defining evaluation methodologies lies in the considered channel model, and its analysis, and simulation tools.

Modeling how the physical channel impacts the transmitted signals has two main objectives for communication engineers.
The first one is to enhance our understanding of the physics of communication systems.
Knowledge of the propagation channel's behavior is critical in order to efficiently design future communication schemes.
This is even more prominent in a wireless context where the propagation medium shows high variance in time, frequency and space.
To handle this, the preferred approach is to use stochastic approximations of the behavior of carrier waves.
How good these approximations are will determine how well the designed technology will work in practice.
This leads us to the second objective of channel modeling: to evaluate the performance of a given technological solution in a realistic environment. 
Environmental complexity requires some simplifications in order to obtain a tractable tool which is used to simulate a very large number of links simultaneously.
Models fulfilling the first goal, namely \emph{design} models, will usually abstract away some of the parameters considered to focus on specific properties of the propagation medium.
On the other hand, the \emph{simulation} models need to cover many propagation environments and technologies.
They are, thus, more complex than their design counterparts, and can be tweaked and configured to handle a larger number of scenarios.

In this article, we highlight the new challenges and requirements that 5G imposes on channel modeling and channel simulation tools \cite{Osseiran2014}. In particular, we analyze how the massive increase of antennas, the move to higher bands and the increasing heterogeneity of cellular deployments are moving the frontier of current modeling technology \cite{Clerckx2012, Jamsa2015}.
We provide a unified view of previous analysis and tutorials, in mobile radio access channels \cite{Jamsa2015} and more specific vehicular and machine to machine (M2M) channels \cite{Viriyasitavat2015, Tlaha2011}.
We also address current topics and trends such as mmWave \cite{Gustafson2014a}, massive MIMO (M-MIMO) channels \cite{Larsson2014,Wu2015}, 3D channels \cite{Cheng2014}, map-based modeling \cite{Jamsa2015,Kasparick2015}, and non-stationary channels \cite{Ghazal2015,Wu2015}.
We review classical models and background information in Sec.~\ref{sec:simul}, and analyze the key drivers in wireless technology that will shape the future of channel modeling in Sec.~\ref{sec:drivers}. This allows us to identify the broad research trends to pursue in the standardization process in Sec.~\ref{sec:trends}. We conclude on a discussion about the challenges ahead.

\section{Simulating wireless channels}\label{sec:simul}
\label{sec:modeling}

Building new models for the propagation channels must start by the definition of the specific scenarios of interest.
A scenario is defined by a \emph{typical} usage of the channel, in a typical environment, with the goal of answering a specific question.
As a bridge between the models used and the reality, great care is usually placed in defining the scenarios of interest in the standardization process; their definition is subject to much debate among the actors in standard bodies.

Simulation models then strive to reproduce the behavior of the chosen scenario, as realistically as possible.
However, there is an inherent trade-off made in practical channel simulation.
For example, one could always model the environment exactly and solve Maxwell's equations.
Notwithstanding the computational complexity of this approach, it would give a very accurate answer at the cost of parametric complexity.
It would also lack generality, in the sense that multiple instances of the environment would have to be averaged over to obtain a typical behavior of the channel in the scenario of interest.
We can analyze most models under these terms, and consider how they trade-off in terms of:
\begin{itemize}
	\item Accuracy, i.e. \emph{how close is the realization to a specific instance of my scenario?}
	\item Generality, i.e. \emph{how typical is the realization with respect to my scenario?}
	\item Simplicity, i.e. \emph{how complex is it to parametrize and run my simulation?}
\end{itemize}
These characteristics are somewhat antagonistic when considering abstract scenarios.
One intuitively expects models with a large number of parameters to be of less generality.
In essence, solving Maxwell's equation exactly, or in a simplified manner through \emph{ray-tracing} would score perfectly if the scenario of interest was limited to a specific environment.
The map-based model of \cite{Jamsa2015} is an example of such a ray-tracer; an even more extreme case is using measurements as a simulation platform, as e.g. in \cite{Adhikary2014}.
Overall, most models will favor some objective rather than others.
Design models have a high emphasis on simplicity, preferring lower numbers of parameters capturing the essence of the model, while
ray-tracer types tend to value accuracy rather than generality. 

Most simulation models stem from simplifications of the ray launching paradigm \cite{Clerckx2012}.
One way to do so is to concentrate on the endpoints, or \emph{drops}, which are an abstraction for the users locations in a large network.
The interacting objects in the environment, also called the \emph{scatterers}, are then randomly generated, parametrized and associated with each drop, to form a statistically representative channel.
This \emph{user-centric} approach is the one chosen by the 3GPP model in its first iteration, the Spatial Channel Model (SCM) and its derivatives.
Another approach is to consider a \emph{scatterer-centric} approach, where scatterers are generated globally so as to form a virtual environment.
Every user then shares the scatterers, mimicking a ray-tracing with a finite number of known interacting objects.
This is the preferred approach of the Directional Channel Model (DCM) which culminated in the COST2100 channel.
All these models and related approaches are described in more details in \cite{Clerckx2012, Jamsa2015} and references therein.

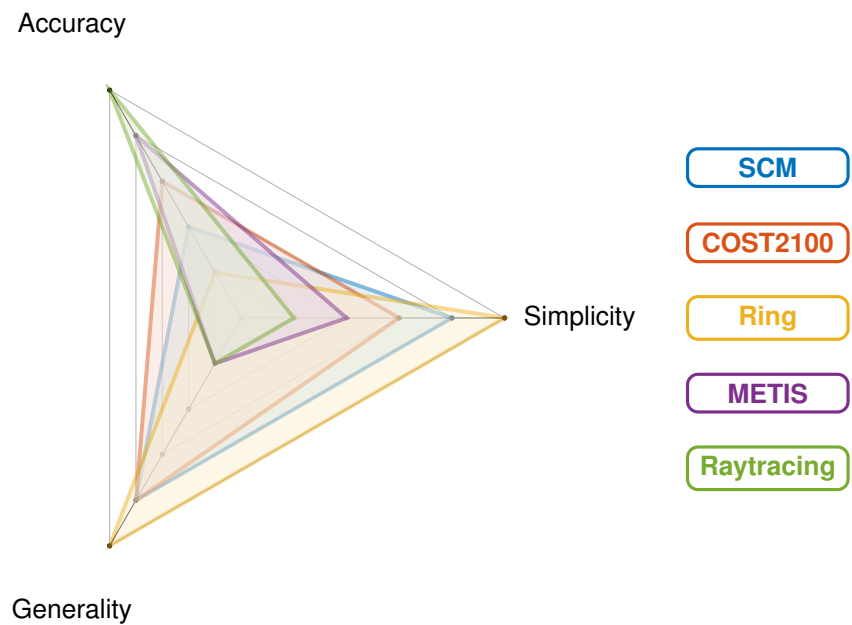
\begin{figure}[p]
	\centering
	% -*- root: wcom-cm.tex -*-

	\newcommand{\D}{3}
	\newcommand{\U}{5}
	\newdimen\R
	\R=3.5cm 
	\newdimen\L
	\L=4.5cm
	\newcommand{\A}{360/\D}
	\begin{tikzpicture}
		\path (0:0cm) coordinate (O);
		\foreach \X in {1,...,\D}{
		    \draw (\X*\A:0) -- (\X*\A:\R);
		  }

		  \foreach \Y in {0,...,\U}{
		    \foreach \X in {1,...,\D}{
		      \path (\X*\A:\Y*\R/\U) coordinate (D\X-\Y);
		      \fill (D\X-\Y) circle (1pt);
		    }
		    \draw [opacity=0.3] (0:\Y*\R/\U) \foreach \X in {1,...,\D}{
		        -- (\X*\A:\Y*\R/\U)
		    } -- cycle;
		  }
		\path (1*\A:\L) node (L1) {\sffamily Accuracy};
		\path (2*\A:\L) node (L2) {\sffamily Generality};
		\path (3*\A:\L) node (L3) {\sffamily Simplicity};
		\draw [color=col1, fill=col1!20,line width=1.5pt,opacity=0.5]
		    (D1-2) --
		    (D2-4) --
		    (D3-4) -- cycle;
		\draw [color=col2, fill=col2!20,line width=1.5pt,opacity=0.5]
		    (D1-3) --
		    (D2-4) --
		    (D3-3) -- cycle;
		\draw [color=col3, fill=col3!20,line width=1.5pt,opacity=0.5]
		    (D1-1) --
		    (D2-5) --
		    (D3-5) -- cycle;
		\draw [color=col4, fill=col4!20,line width=1.5pt,opacity=0.5]
		    (D1-4) --
		    (D2-1) --
		    (D3-2) -- cycle;
		\draw [color=col5, fill=col5!20,line width=1.5pt,opacity=0.5]
		    (D1-5) --
		    (D2-1) --
		    (D3-1) -- cycle;
		\node[text width=1.9cm, very thick, draw, rounded corners, align=center, col1] at (7, 2) {\sffamily \textbf{SCM}};
		\node[text width=1.9cm, very thick, draw, rounded corners, align=center, col2] at (7, 1) {\sffamily \textbf{COST2100}};
		\node[text width=1.9cm, very thick, draw, rounded corners, align=center, col3] at (7, 0) {\sffamily \textbf{Ring}};
		\node[text width=1.9cm, very thick, draw, rounded corners, align=center, col4] at (7, -1) {\sffamily \textbf{METIS}};
		\node[text width=1.9cm, very thick, draw, rounded corners, align=center, col5] at (7, -2) {\sffamily \textbf{Raytracing}};

	\end{tikzpicture}
	\caption{Categorization of the tradeoffs made by different modern simulation and design models, used for cellular networks research and standardization. The SCM and its family (SCME, WINNER, Quadriga, ...) are derived from the standardized 3GPP model. The COST model family and ring models at large are discussed in \cite{Clerckx2012} and references therein, as well as in \cite{Jamsa2015}. The METIS map-based raytracing model specifications can be found in \cite{Jamsa2015}.}
	\label{fig:spider}
\end{figure}

We illustrate the relative strengths of each model families according to the three aforementioned dimensions of accuracy, simplicity and generality in Fig.~\ref{fig:spider}.
This allows to quickly assess how they trade off the different objectives and, thereby, how one may choose a model depending on specific objectives.
While the latest iteration of these models cover the current standardization needs, further developments are required to account for the requirements of future communication systems \cite{Osseiran2014}.
As the use cases of the network evolves, the scenarios of interest evolve with them, and simulation models need to cater to these new simulation needs.
Identifying the key drivers behind future technology is therefore of paramount importance in order to efficiently work toward the next iteration of channel models.

\section{Key drivers for new channel models}\label{sec:drivers}

There still is much debate over the specifications of the future mobile networks and the associated technologies \cite{Osseiran2014}.
While detailing them is beyond the scope of the paper, at their core is a need for higher spectral efficiency, and a move toward more varied usage scenarios of interest in an increasingly heterogeneous network (HetNet) infrastructure.
This, in turn, leads to a number of specific technological paths that we discuss now.

\subsection{Large antenna arrays}

One way to dramatically increase spectral efficiency is by using multiple antennas simultaneously at the transmitter and/or receiver side, in order to exploit multi-path propagation and the inherent \emph{multi-user} diversity of the wireless channel \cite{Clerckx2012}. During the last decade, multiple antennas have been integrated on devices and base stations (BSs), in order to achieve the theoretical promises of increased capacity.
Multiple-input multiple-output (MIMO) has become an essential element of wireless communication and is present in many standards including but not limited to IEEE 802.11n/ac, IEEE 802.16d, and the 3GPP standard family related to Long Term Evolution (LTE).
The technology has also been extended to improve both the robustness and the performance of single links, as well as a multiple-access method to simultaneously serve users separated in space on the same frequency band.

The trend now goes in the direction of massively increasing the number of antennas, which opens a number of new technological possibilities \cite{Larsson2014}.
In a way, massive MIMO (M-MIMO) gives the transmitter much more freedom in designing its output energy pattern.
The large number of antennas enables very precise \emph{beamforming}, targeting users without creating interference, or selectively removing interference from some points in space.
The increase of the number of antennas can either take the form of co-located antennas placed on wider panels such as walls or by aggregating distributed antennas from various sites, leading to so-called distributed MIMO (D-MIMO). 

\begin{figure}[p]
	\centering
	% -*- root: wcom-cm.tex -*-

\begin{tikzpicture}[y={(-1cm,0.5cm)},x={(1cm,0.5cm)}, z={(0cm,1cm)}, scale=2]
	\newcommand{\beamf}[8]{
		% \pgfmathsetmacro{\xo}{#1}
		% \pgfmathsetmacro{\xt}{#2}
		% \pgfmathsetmacro{\yo}{#3}
		% \pgfmathsetmacro{\yt}{#4}
		% \pgfmathsetmacro{\px}{#5}
		% \pgfmathsetmacro{\py}{#6}
		% \pgfmathsetmacro{\pz}{#7}
		\draw[fill=#8!50, #8, opacity=0.5] (#1, #3, 0) -- (#1, #4, 0) -- (#2, #4, 0) -- (#2, #3, 0) -- cycle;
		\draw[fill=#8!50, #8, opacity=0.5] (#5, #6, #7) -- (#1, #3, 0) -- (#1, #4, 0) -- cycle;
		\draw[fill=#8!50, #8, opacity=0.5] (#5, #6, #7) -- (#1, #4, 0) -- (#2, #4, 0) -- cycle;
		\draw[fill=#8!50, #8, opacity=0.5] (#5, #6, #7) -- (#2, #3, 0);
	}
	\begin{scope}[yshift=1.25cm, xshift=-2.5cm]
		\draw[black!80, fill=black!10] (0, -1, 0) -- (0, 1, 0) -- (0, 1, 2) -- (0, -1, 2) -- cycle;	
		\draw[black!80, fill=black!30] (0, -1, 2) -- (2, -1, 2) -- (2, 1, 2) -- (0, 1, 2) -- cycle;	
		\draw[black!80, fill=black!20] (0, -1, 0) -- (2, -1, 0) -- (2, -1, 2) -- (0, -1, 2) -- cycle;
		\foreach \f in {0.85, 0.55, 0.25, -0.05, -0.35, -0.65}
		{
			\draw[black!80, fill=white] (0, \f, 0.3)	-- (0, \f-0.2, 0.3) -- (0, \f-0.2, 0.7) -- (0, \f, 0.7)-- cycle;
			\draw[black!80, fill=white] (0, \f, 0.8)	-- (0, \f-0.2, 0.8) -- (0, \f-0.2, 1.2) -- (0, \f, 1.2)-- cycle;
			\draw[black!80, fill=white] (0, \f, 1.3)	-- (0, \f-0.2, 1.3) -- (0, \f-0.2, 1.6) -- (0, \f, 1.6)-- cycle;
		}
		\begin{scope}[canvas is yz plane at x=0]
			\draw[col1] (-0.4, 1.9) -- (0.4,1.9);
			\foreach \c in {-0.4,-0.3,..., 0.4}
			{\draw[col1, fill=col1] (\c, 1.9) circle (0.02);}
			\node[col1] at (-0.3, 2.6) {\sffamily Horizontal array};
			\draw[col1, -latex]   (-0.3, 2.5) -- (0, 1.9);
		\end{scope}
		\beamf{-1}{-0.1}{-0.4}{-0.6}{0}{-0.05}{1.9}{col1}
		\beamf{-1}{-0.1}{0.6}{0.4}{0}{0.05}{1.9}{col1}
		\beamf{-1}{-0.1}{0.07}{-0.07}{0}{0}{1.9}{col1}
		\beamf{-1}{-0.1}{-0.9}{-1.2}{0}{-0.1}{1.9}{col1}
		\beamf{-1}{-0.1}{1.2}{0.9}{0}{0.1}{1.9}{col1}
	\end{scope}
	\draw[black!80, fill=black!10] (0, -1, 0) -- (0, 1, 0) -- (0, 1, 2) -- (0, -1, 2) -- cycle;	
	\draw[black!80, fill=black!30] (0, -1, 2) -- (2, -1, 2) -- (2, 1, 2) -- (0, 1, 2) -- cycle;	
	\draw[black!80, fill=black!20] (0, -1, 0) -- (2, -1, 0) -- (2, -1, 2) -- (0, -1, 2) -- cycle;
	\foreach \f in {0.85, 0.55, 0.25, -0.05, -0.35, -0.65}
	{
		\draw[black!80, fill=white] (0, \f, 0.3)	-- (0, \f-0.2, 0.3) -- (0, \f-0.2, 0.7) -- (0, \f, 0.7)-- cycle;
		\draw[black!80, fill=white] (0, \f, 0.8)	-- (0, \f-0.2, 0.8) -- (0, \f-0.2, 1.2) -- (0, \f, 1.2)-- cycle;
		\draw[black!80, fill=white] (0, \f, 1.3)	-- (0, \f-0.2, 1.3) -- (0, \f-0.2, 1.6) -- (0, \f, 1.6)-- cycle;
	}
	\begin{scope}[canvas is yz plane at x=0]
		\draw[col2] (0, 1) -- (0,1.7);
		\foreach \c in {1,1.1,...,1.8}
		{\draw[col2, fill=col2] (0, \c) circle (0.02);}
		\node[col2] at (0, 2.4) {\sffamily Vertical array};
		\draw[col2, -latex]   (0, 2.3) -- (0, 1.7);
	\end{scope}
	\beamf{-0.8}{-0.5}{1}{-1}{0}{0}{1.3}{col2}
	\beamf{-1.5}{-1}{1}{-1}{0}{0}{1.4}{col2}
	\begin{scope}[canvas is xz plane at y=-1]
		\foreach \x in {0.6,0.7,...,1.5}
		{
			\draw[col4] (\x, 1) -- (\x, 1.7);
			\foreach \z in {1,1.1,...,1.8}
			{
				\draw[col4, fill=col4] (\x, \z) circle (0.02);
			}
		}
		\node[col4] at (0.6, 2.6) {\sffamily Rectangular array};
		\draw[col4, -latex]   (0.6, 2.5) -- (1, 1.7);
	\end{scope}
	\beamf{0.3}{0.6}{-1.2}{-1.5}{1}{-1}{1.4}{col4}
	\beamf{0.5}{0.7}{-1.7}{-2}{1}{-1}{1.4}{col4}
	\beamf{1.1}{1.4}{-1.5}{-1.7}{1}{-1}{1.4}{col4}
	\beamf{1.4}{1.8}{-2}{-2.4}{1}{-1}{1.4}{col4}
\end{tikzpicture}
	\caption{Full dimension beamforming is a technique that uses the actual shape of an antenna array to direct beams in specific directions. We see here that along the direction of the array, we can create strong beams that targets specific zones, and thereby multiplex users in space. Simulating such a technique requires a precise parametrization of the environment in every dimensions.}
	\label{fig:3Dmod}
\end{figure}
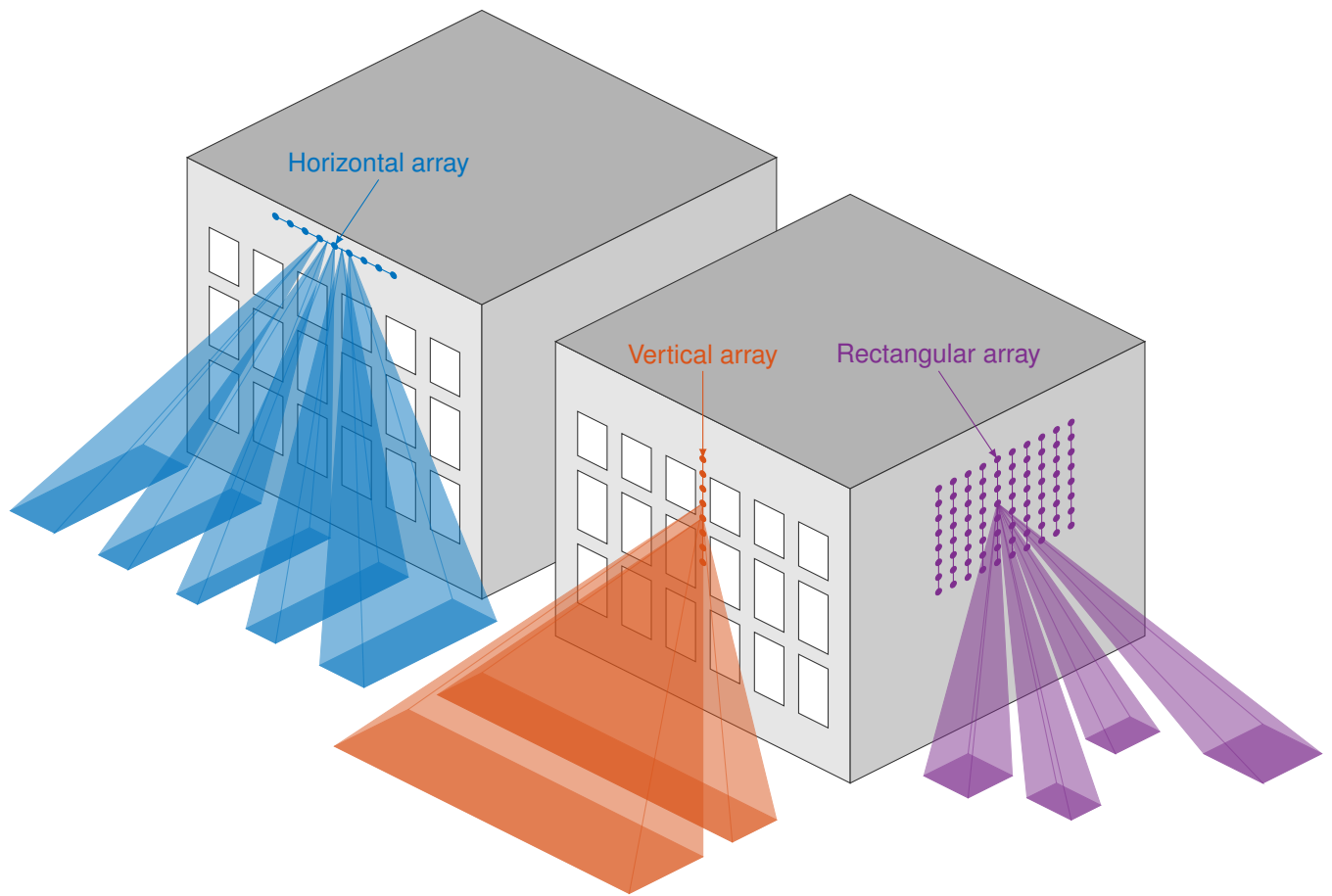

A very specific need for MIMO is related to the beamforming and the considered antenna patterns.
While many cellular simulations were limited to a 2-D plane with linear antenna arrays, advanced MIMO communications require a complete description of the space.
Adding the elevation dimension was an intricate task, which required model adaptation as well as new measurements.
Most of the current simulation models have integrated this 3-D modeling, for both transmitting and receiving antennas, thereby enabling the evaluation of more complex antennas structure and scenarios (see e.g. Fig.~\ref{fig:3Dmod}).
Full dimensional assumptions are now the norm, and are integrated into the latest iterations of the major simulation models \cite{Cheng2014}.
However, the parametrization and calibration of these models for different scenarios is still an ongoing matter.

\subsection{New frequency bands}

Another approach for increasing the wireless capacity is by reaching toward new frequency bands with a lot of available bandwidth. 
Advances in hardware have enabled communication scientists to consider alternative bands out of the overloaded conventional licensed ones.
Frequencies below 6 GHz are more and more flooded by communication systems with low quality of service (QoS) guarantees such as WiFi and Bluetooth.
They also handle most of the land-to-land mobile communications today.
As such, the behavior of carrier waves in this frequency range is now well understood.

Beyond 6 GHz, multiple bands are starting to be considered for future communication systems such as the 28 GHz and the 73 GHz band, avoiding the high absorption effect of oxygen at 60 GHz \cite{Gustafson2014a}.
These frequencies are in the process of being decommissioned from their earlier uses, if any, and re-licensed for mobile communications worldwide.
Since they were never used for ground-to-ground communication, their propagation characteristics in radio access scenarios are not yet well known.
Being free of regulations and used in the IEEE 802.11ad standard, the 60 GHz band is an exception and has been extensively evaluated in indoor scenarios.
A large body of work has thus improved our understanding of these bands, although multiplexing capabilities and multipath behavior is still under study (e.g. \cite{Gustafson2014a,Adhikary2014} and references therein).
All these elements in addition to the standardization efforts have motivated researchers to launch measurement campaigns aiming at better understanding how carrier waves behave at mmWave frequencies, and how different is this behavior compared to lower bands. 

\subsection{New deployments}

In addition to the above requirements related to new technologies that came as a response to the promised thousand-fold throughput improvement \cite{Osseiran2014}, there has been a lot of effort in proposing new network architectures.
The idea of using smaller and denser cells in particular has emerged to the forefront, after being discarded in the early days of cellular networks due to the complexity and cost of their deployment.
Access points getting closer to the user means that the network is more dynamic in nature, with frequent switches between serving BSs from the user point of view.
A user could also be served jointly by multiple BSs, a possibility that is already enabled in the more recent LTE standard through the so-called Coordinated Multipoint (CoMP) schemes.
Moreover, the emergence and exponential increase of connected objects ranging from vehicles to wearables brings in a completely new kind of propagation channels to be investigated and modeled.
This trend is commonly termed at the Internet of Things (IoT).
In the future, we also expect that direct communication between users and objects, in a device-to-device (D2D) manner, might then be used either to offload some part of the traffic in the network or to improve coverage. 
Such extensions are already studied in the current 3GPP standards as Proximity Services (ProSe).
Future 5\textsuperscript{th} generation (5G) networks, such as IMT-2020, will thus most certainly use a wide variety of link types and technologies in order to achieve the performance and flexibility that are expected from them \cite{Osseiran2014}.
		
All these new techniques and network deployments differ wildly from the classical architectures from a channel modeling point of view.
For example in a macro cellular network, the BSs are at elevations of several tens of meters.
On the other hand, micro and femto cells would be at heights of a couple of meters.
This difference of perspective changes dramatically the nature of the propagation channel, the obstacles and their nature as well as their density and distribution.
This implies that the conventional channel models defined for a classical cellular system are no longer valid and must be updated or replaced if necessary.
In addition to these propagation constraints, we notice that for some new scenarios of interest, both endpoint are possibly moving objects \cite{Viriyasitavat2015}.
In fact, for all scenarios where mobile relays or D2D communication is considered, both the transmitter and the receiver are mobile in \emph{random} directions \cite{Tlaha2011}.
In cluttered environments, this implies that their channel will be fast-varying, and require the inclusion of new dynamics in their modeling \cite{Hlawatsch2011}.
There have been adaptations of the classical channel model families to the needs of D2D, e.g. the SCM/WINNER model in the ProSe discussions in 3GPP, but they are usually limited to adaptation of basic large-scale parameters.
The more fine-grained behavior and especially the joint behavior of radio links are still not the focus of current model architectures.

\section{New trends in channel modeling}\label{sec:trends}

These new scenarios highlight both deficiencies and opportunities in the channel models and simulation tools currently in use.
On one hand, they make a number of simplifying assumptions, which have been validated with respect to their original application scenarios.
When the application scenarios change or evolve, those simplifications may not hold anymore, and the fundamental model needs to be adapted or deeply revised.
On the other hand, analyzing the physical behavior of wireless propagation channels may lead to new opportunities for system engineers.
Specific behaviors can be harnessed through advanced algorithms, as enablers for new communication algorithms.
One common example is the 2-level precoding method described in \cite{Adhikary2014}.

\subsection{Higher spatial resolution}

The consideration of elevation -- the third dimension -- in channel models has already prepared the technical tools to evaluate advanced beamforming \cite{Cheng2014}.
A very promising aspect of elevation beamforming is for smaller cells with large antenna arrays with respect to the user distance.
Consider a practical case of such a small cell on the side of a building in a street canyon, as exemplified in Fig.~\ref{fig:3Dmod}.
Using a comparatively low number of antennas in a vertical array, the smaller cell can then be expected to discriminate its users in space, and achieve large multi-user MIMO gains.
While this conjecture has been validated in practice e.g., in \cite{Cheng2014}, one remaining question is on how to model the channel seen through these larger antenna arrays.
In current models, there is a basic assumption that the antenna aperture is small with respect to the distance between the transmitter, the receiver, and its scatterers.
Consequently, they may be considered as points and the wavefronts on all antennas can be modeled as parallel planes.
This has a tremendous impact on the complexity of simulations, as all antennas will see similar paths from the transmitter to the receiver.
The path for one antenna pair can be computed and subsequently replicated over the array, accounting for the small spatial displacement between neighboring antennas.
This hypothesis is realistic for macro-cellular BSs with co-located antennas and distant users.
However, in novel uses of the channel, this breaks down in a number of ways.

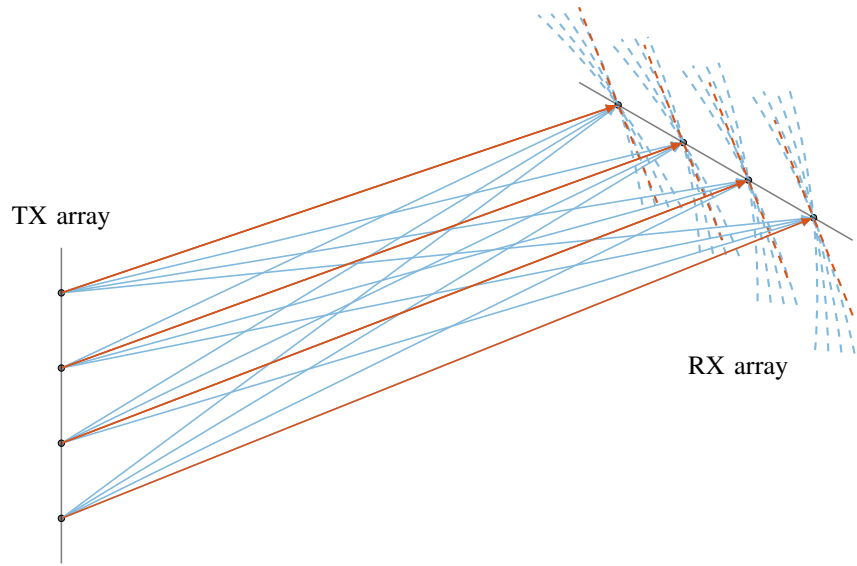
\begin{figure}[p]
	\centering
	% -*- root: wcom-cm.tex -*-

\begin{tikzpicture}
\tkzInit[xmin=-2,xmax=10,ymin=-1,ymax=10]
\tkzDefPoint(0,0){T1}
\tkzDefPoint(0,1){T2}
\tkzDefPoint(0,2){T3}
\tkzDefPoint(0,3){T4}
\tkzDrawPoints(T1,T2,T3,T4)
\tkzDefPoint(10,4){R1}
\tkzDefShiftPoint[R1](150:1){R2}
\tkzDefShiftPoint[R1](150:2){R3}
\tkzDefShiftPoint[R1](150:3){R4}
\tkzDrawPoints(R1,R2,R3,R4)
\tkzDefPoint(0,1.5){T0}
\tkzDefShiftPoint[R1](150:1.5){R0}
\tkzDefPointWith[orthogonal normed](R1,T1) \tkzGetPoint{Pl11}
\tkzDefPointWith[orthogonal normed, K=-1](R1,T1) \tkzGetPoint{Pl12}
\tkzCompass[delta=10, dashed, thick, color=col1!50](T1,R1)
\tkzCompass[delta=10, dashed, thick, color=col1!50](T2,R1)
\tkzCompass[delta=10, dashed, thick, color=col1!50](T3,R1)
\tkzCompass[delta=10, dashed, thick, color=col1!50](T4,R1)
\tkzCompass[delta=10, dashed, thick, color=col1!50](T1,R2)
\tkzCompass[delta=10, dashed, thick, color=col1!50](T2,R2)
\tkzCompass[delta=10, dashed, thick, color=col1!50](T3,R2)
\tkzCompass[delta=10, dashed, thick, color=col1!50](T4,R2)
\tkzCompass[delta=10, dashed, thick, color=col1!50](T1,R3)
\tkzCompass[delta=10, dashed, thick, color=col1!50](T2,R3)
\tkzCompass[delta=10, dashed, thick, color=col1!50](T3,R3)
\tkzCompass[delta=10, dashed, thick, color=col1!50](T4,R3)
\tkzCompass[delta=10, dashed, thick, color=col1!50](T1,R4)
\tkzCompass[delta=10, dashed, thick, color=col1!50](T2,R4)
\tkzCompass[delta=10, dashed, thick, color=col1!50](T3,R4)
\tkzCompass[delta=10, dashed, thick, color=col1!50](T4,R4)
\tkzDefShiftPoint[Pl11](150:1){Pl21}
\tkzDefShiftPoint[Pl11](150:2){Pl31}
\tkzDefShiftPoint[Pl11](150:3){Pl41}
\tkzDefShiftPoint[Pl12](150:1){Pl22}
\tkzDefShiftPoint[Pl12](150:2){Pl32}
\tkzDefShiftPoint[Pl12](150:3){Pl42}
\tkzDrawLine[dashed, color=col2, thick](Pl11,Pl12)
\tkzDrawLine[dashed, color=col2, thick](Pl21,Pl22)
\tkzDrawLine[dashed, color=col2, thick](Pl31,Pl32)
\tkzDrawLine[dashed, color=col2, thick](Pl41,Pl42)
% \tkzDrawLine[dashed](Pl41,Pl42)
\tkzDrawSegment[-latex, color=col1!50](T1,R1)
\tkzDrawSegment[-latex, color=col1!50](T1,R4)
\tkzDrawSegment[-latex, color=col1!50](T1,R3)
\tkzDrawSegment[-latex, color=col1!50](T1,R2)
\tkzDrawSegment[-latex, color=col1!50](T2,R1)
\tkzDrawSegment[-latex, color=col1!50](T2,R4)
\tkzDrawSegment[-latex, color=col1!50](T2,R3)
\tkzDrawSegment[-latex, color=col1!50](T2,R2)
\tkzDrawSegment[-latex, color=col1!50](T3,R1)
\tkzDrawSegment[-latex, color=col1!50](T3,R4)
\tkzDrawSegment[-latex, color=col1!50](T3,R3)
\tkzDrawSegment[-latex, color=col1!50](T3,R2)
\tkzDrawSegment[-latex, color=col1!50](T4,R1)
\tkzDrawSegment[-latex, color=col1!50](T4,R4)
\tkzDrawSegment[-latex, color=col1!50](T4,R3)
\tkzDrawSegment[-latex, color=col1!50](T4,R2)
\tkzDrawSegment[-latex, color=col2](T1,R1)
\tkzDrawSegment[-latex, color=col2](T2,R2)
\tkzDrawSegment[-latex, color=col2](T3,R3)
\tkzDrawSegment[-latex, color=col2](T4,R4)
\tkzDrawSegment[-latex, color=col2](T2,R2)
\tkzDrawSegment[-latex, color=col2](T3,R3)
\tkzDrawSegment[-latex, color=col2](T4,R4)
\tkzDrawLine[opacity=0.5](T1,T4)
\tkzDrawLine[opacity=0.5](R1,R4)
\node at (0, 4) {TX array};
\node at (9, 2) {RX array};
\end{tikzpicture}
	\caption{Propagation between two antennas arrays, comparing the true circular wavefronts (in blue) and the classical planar wavefront approximation (in red). For the planar approximation, the distance between the antennas -- and thus the phase offset -- depends only on the antenna angle with respect to the impinging angle of the wave, and the receiving antenna separation. When the wavefront is not planar however, the phase offset has to be computed from the distance between antenna pairs. This increases the diversity in the channel and thus improves the overall performance.}
	\label{fig:spherical}
\end{figure}

A key difference lies in the stationarity of the channel model over the antenna array, an issue we discuss in Sec.\ref{sec:nonstat}.
Another difference is that, as the array size increases, cell coverage gets smaller and BSs move closer to their users, thereby, crossing the Rayleigh distance.
Consequently, propagation characteristics change from the far-field to the radiative near-field, and the wavefront can not reliably be assumed to be planar anymore \cite{Jiang2005}.
Note that one can still approximate the general solution of Maxwell's equation by the ray-launching paradigm; in this case, however, the phase offset of the wave between adjacent antennas depends on the exact distance between transmit and receive antenna pairs, rather that the receiver antenna separation and angle of arrival of the wave (Fig.~\ref{fig:spherical}).
While this will increase the computational complexity of the model, this also comes with a positive aspect.
Line-of-sight MIMO communications with a low number of multi-path components are not expected to provide much gains in the far-field.
However, the authors of \cite{Jiang2005} have shown that in the radiative near-field, this planar approximation can actually cause the capacity to be severely underestimated for close range communications. 

\subsection{Joint behavior across time, space and frequency}

The move to higher bands is only one facet of the search for spectral efficiency.
LTE Releases 12 and 13 already contain multiple work items and standardization points on carrier aggregation -- a technique that aims at jointly transmitting and receiving on multiple frequency bands.
For some of these bands, the marginal behavior of the propagation environment is well known.
For higher bands, it is in the process of being assessed using conventional methods \cite{Jamsa2015,Gustafson2014a}.
Meanwhile, a central question remains: how to correctly measure and model the joint behavior of links in channel simulations?

The modeling of correlation in space is partially present in many simulation tools, especially for shadowing and masking effects of relatively large scale, for example the presence of buildings or massive objects between transceivers.
Receivers that are geographically co-located will have a global pathloss that is correlated.
The treatment of correlation at the scatterer-level on the other hand varies between the approaches and model families.
In scatterer-centric modeling, scatterers can naturally be shared between different users or linked together as \emph{twin-clusters} \cite{Clerckx2012}.
The evolution of the channel state on short time and spatial scales is, thus, handled by the model.
User-centric models on the other hand have limited this joint modeling to large-scale parameters such as shadowing \cite{Jamsa2015}.
This approach has a clear advantage from a complexity point of view.
However, it fails to capture the finer aspects in the joint modeling of propagation links, which future technologies are expected to exploit.
Remote antennas, and D-MIMO in general, are in part reliant on this joint evolution of the link to provide capacity gains.
However, one risks overestimating the gains of cooperation if the joint distribution of the links is not taken into account.
Few simulation tools are able to properly assess this, as even ray-tracers will usually not correlate the phases of similar paths to the destination \cite{Jamsa2015}.

This need for joint modeling extends to the time and frequency domains. 
While we discuss the former in Sec.\ref{sec:nonstat} on non-stationary models, the joint behavior over frequency bands on the other hand remains largely unknown.
Very few works and fewer simulation tools consider it, although it is a key enabler for many potential technologies -- exemplified by the heavy focus of current standardization efforts on carrier aggregation between classical bands.
Due to the specific and very directional nature of the channel in the mmWave bands, designers expect to use classical bands for discovery, control and broadcast tasks, while the huge bandwidth available in the high bands will service data transfer \cite{Osseiran2014}.
If one can infer the channel state of the high bands from the channel state of the lower bands, such aggregation is much simplified.
This relates in general to frequency-division duplex (FDD) and channel state information.
The lack of reciprocity of the channel between frequency bands is, at large, a significant problem for the future generation of wireless.
Massive MIMO for example can not function properly in FDD due to this phenomenon \cite{Larsson2014}.
Evaluating approaches trying to alleviate this problem, such as \cite{Decurninge2015}, will require proper models at the system level that are not yet available.

\subsection{Non-stationary models}
\label{sec:nonstat}

A common assumption in most channel models is that their distributions are stable in time, frequency and space.
In particular, it is often expected that the autocorrelation of the channel impulse response is wide-sense stationary (WSS) over time, and that fading due to the scatterers is uncorrelated and independent between different \emph{drops} in space \cite{Clerckx2012,Hlawatsch2011}.
For example, in the 3GPP SCM, coherence in time is considered only through a Doppler component \cite{Clerckx2012}.
Each user is associated to a speed vector, which translates into Doppler effects for each sub-path from each scatterer.
The underlying approximation here is that a moving user will see a similar environment during its displacement, so that large- and small-scale parameters are fixed during the simulation of a specific drop.
How large this displacement can be in time or frequency can be analyzed and estimated through measurements \cite{Hlawatsch2011}.
It may also be analyzed for specific channel models \cite{Wu2015}.
However, in some scenarios of interest, the channel statistics are not constant over the transmission period and scatterers around the transceivers can appear and disappear rapidly.
This is for example the case for D2D in very cluttered areas \cite{Tlaha2011}, or in communication with or between high-speed vehicles such as trains and cars \cite{Viriyasitavat2015, Ghazal2015}.
As we discuss in Sec.~\ref{sec:maps}, parameter maps can handle a part of this problem from the channel simulation side by correlating large-scale parameters over space and time.
Specific simulation models have also been proposed to simulate fast-varying non-stationary channels, with an adequate parametrization \cite{Ghazal2015}.

For larger M-MIMO arrays, there is strong evidence that all antennas will not see the same interacting objects, or even the same users in extreme cases \cite{Larsson2014}.
This relates to stationarity and homogeneity of scattering effects over an antenna array, in the spatial domain \cite{Clerckx2012}.
While this effect was negligibly small in classical MIMO, it is expected to be of notable importance for M-MIMO.
It opens up new options for user separation and grouping in the spatial dimension, as illustrated, e.g., in \cite{Adhikary2014} for a M-MIMO system at 28 GHz.
In essence, one can use the second order statistic of the spatial channel to extract and group users with common scatterers or similar spatial signatures.
How to analyze and model this form of non-stationarity in space is still open to discussion, while the preferred models are birth and death processes of scatterers in time and space \cite{Wu2015}.
A similar phenomenon appears in D-MIMO scenarios, where neighboring users will share scatterers that evolve jointly over time and space.

\subsection{Data-centric propagation models}
\label{sec:maps}

Heterogeneous cell sizes and deviations from the cellular paradigm, such as D2D, need to be characterized by a plethora of new radio propagation models, each of which comes with a large parameter set.  
It is a dominating trend that these propagation environments are characterized more and more accurately with the help of environmental data.
The 3GPP SCM already defined \emph{maps} of so-called large-scale parameters, in order to capture the joint spatial evolution of characteristics such as shadowing.
New models define very specific scenarios based on building geometry and attenuation material, with the best example being the proposed \emph{map-based} channel model of the METIS project \cite{Jamsa2015}.
This solution seems to be one way to answer the problems discussed up to now with tractable complexity, and support at the same time new modeling constraints, new deployments with dual mobility, mesh networks and non-stationarity in addition to the conventional parameters supported by stochastic modeling.
The model uses multi-dimensional maps of attenuators based on %in function of 
the frequency and positions of the transmitter and receiver, their relative velocities and bandwidth, which are then complemented by randomly placed scatterers and reflectors.

The resulting propagation models are highly diverse, use a large set of data, and often sacrifice generality for accuracy.
But an even more practical approach is to directly use radio propagation measurements for the analysis and design of the radio access.
Such propagation data is widely available, and can be constructed via different means.
One solution would be through extensive simulations by means of ray-tracers.
Since modern handsets provide cost efficient access to measurements of radio signal strength and geographical position, data collection during measurement campaigns during the normal operation could provide another solution.
Operators can access this information via the standardized interfaces e.g. in the latest LTE specifications for the \emph{Minimization of Drive Tests (MDT)}, or through over-the-top service providers such as \url{http://opensignal.com/} and \url{http://www.rootmetrics.com/} who collect and offer such information.

\begin{figure}[p]
	\centering
	\includegraphics[width=1\textwidth]{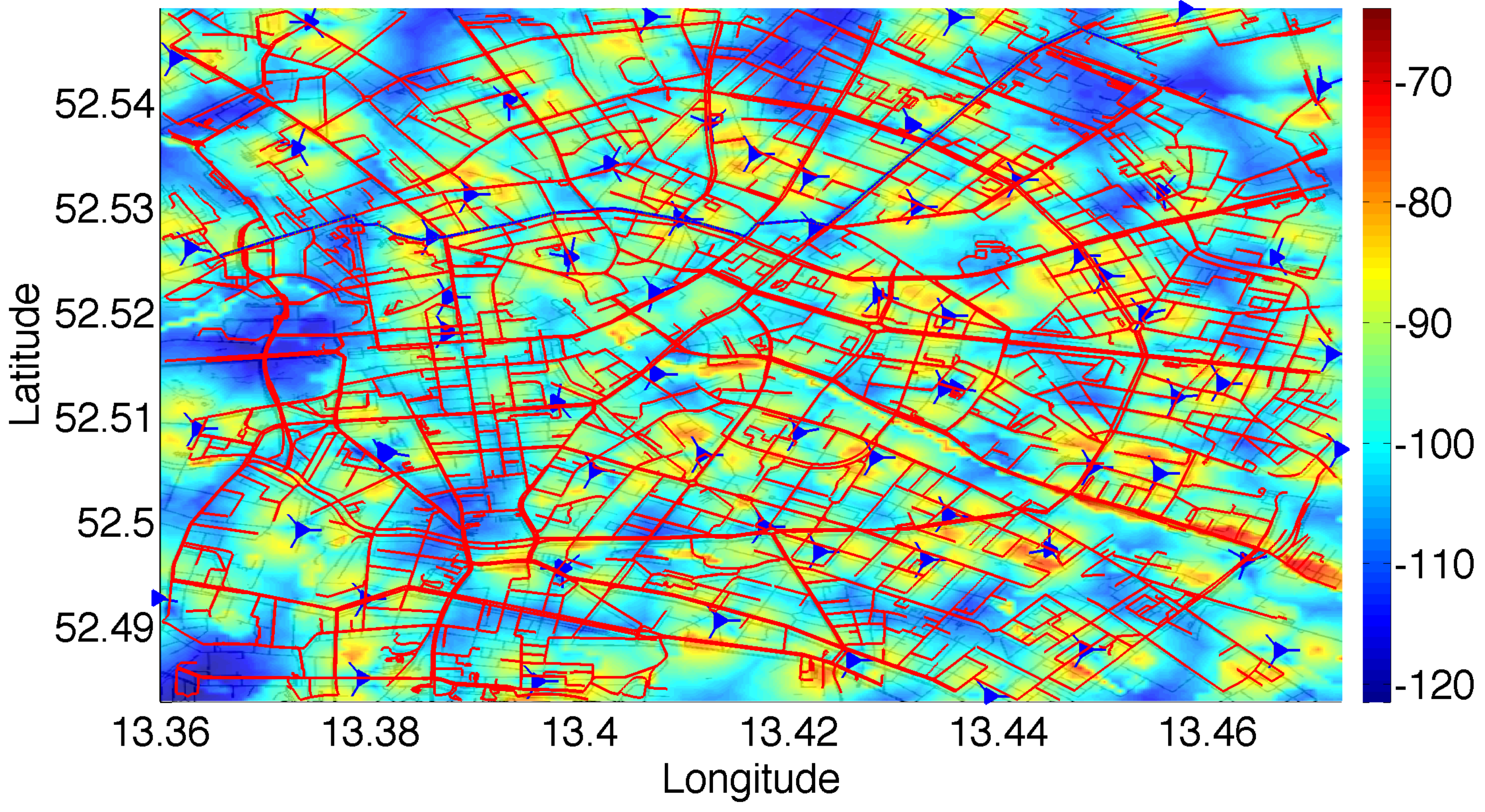}
	\caption{A completed radio map of Berlin, Germany, reconstructed from online measurements \cite{Kasparick2015}: The heatmap shows radio propagation loss between the corresponding position and the strongest serving sector over a geographical area of 56 km$^2$, while the red lines indicate the main roads of the downtown area. Further overlays illustrate building structures in light gray and the stations as blue triangles.}
	\label{fig:map}
\end{figure}

These new, cost-efficient data sources are appealing to model radio propagation based on data alone. However, accurate channel modeling is still necessary, albeit in an indirect hidden manner.
The varying and partly unknown accuracy of the above data sources requires careful processing to detect outliers and to interpolate incomplete information.
As it is often impractical or impossible to measure the channel at every geographical position in a given area, radio propagation maps are often incomplete in terms of area, altitude and sample space.
To fill in the gaps, the community has successfully applied general radio propagation models to parametrize generic methods such as Kriging, matrix completion, and support vector machines \cite{Kasparick2015}.
As illustrated in Fig.~\ref{fig:map}, radio propagation measures may be augmented by further integrating information such as building geometry, street maps, and base station positions, to improve the parametrization and to reduce the search space for machine learning methods.
Such combination of generic models with specific scenario data, provides an interesting trade off between generality and accuracy and is thus a promising field of future research.

\section{Outlook and challenges}

Overall, there is the global need for the more advanced modeling of radio links.
Advanced models need to be more precise, need to cope with the fact that common approximations are not holding anymore for new radio technologies, and need to account for coupled links.
This translates into a higher computational complexity for the simulation tools, and larger sets of parameters to consider for the modeler.
As computational power increases, the former can be handled through better computing equipment and algorithms improving over time.
The latter on the other hand is a burden that has yet to be tackled. 
Radio maps provide an immediate yet incomplete and highly complex answer considering the amount of data that needs to be collected, stored, and processed.
As models get more complex simulations loose generality and, consequently, have to be replicated over varying scenarios to derive general findings.

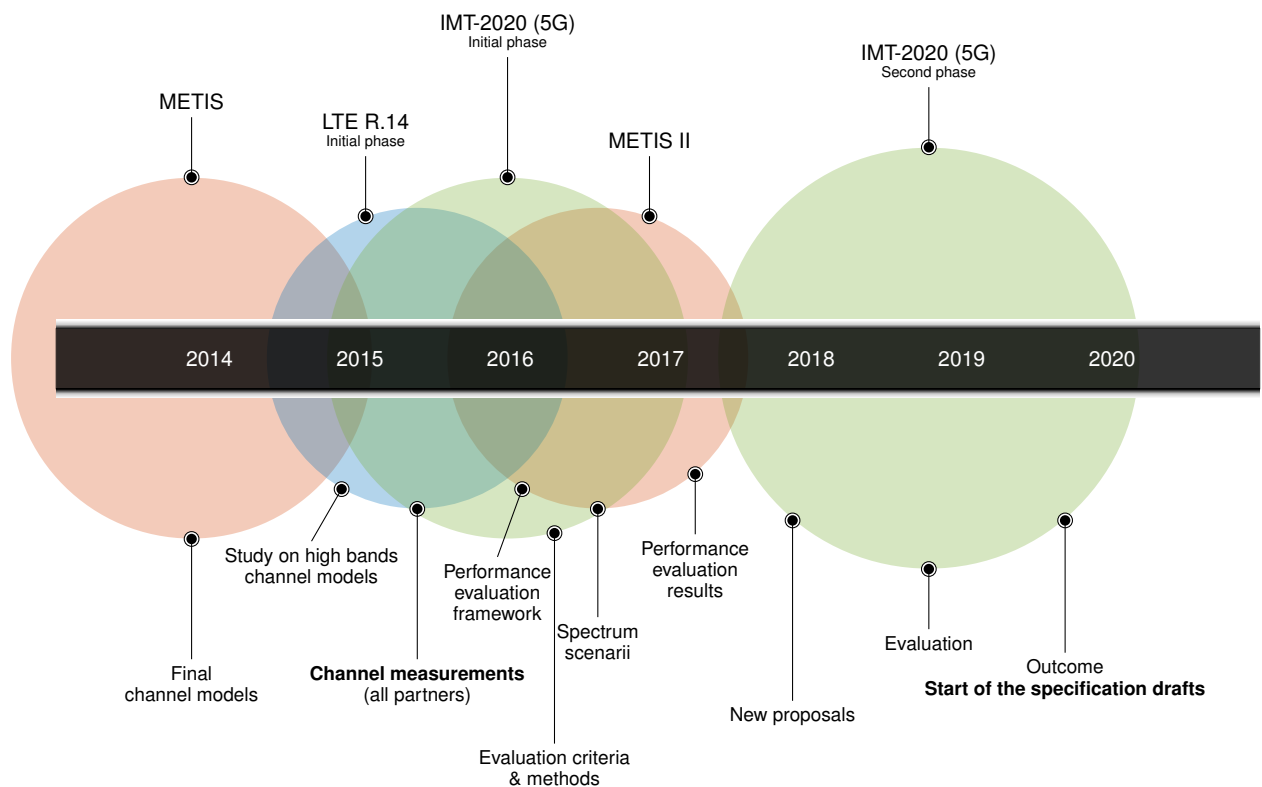
\begin{figure}[p]
	\centering
	% -*- root: wcom-cm.tex -*-

\begin{tikzpicture}[timespan={}, font=\sffamily, transform shape, scale=0.8]
\timeline[custom interval=true]{2014, 2015, 2016, 2017, 2018, 2019, 2020}
  \begin{phases}
    \phase{between week=0 and 1 in 0.9,
    	involvement degree=6cm,phase color=col2}
    \phase{between week=3 and 4 in 0.6,
    	involvement degree=5cm,phase color=col2}
    \phase{between week=2 and 3 in 0.4,
      involvement degree=5cm,phase color=col1}
    \phase{between week=2 and 3 in 1,
      involvement degree=6cm,phase color=col5}
  	\phase{between week=5 and 6 in 0.8,
      involvement degree=7cm,phase color=col5}
  \end{phases}
  \setlength{\baselineskip}{8pt}
  \addmilestone{at=phase-1.90,direction=90:1cm,text={METIS},text options={above, align=center}}
  \addmilestone{at=phase-2.70,direction=90:1cm,text={METIS II},text options={above, align=center}}
  \addmilestone{at=phase-3.110,direction=90:1cm,text={LTE R.14 \\ \scriptsize Initial phase},text options={above, align=center}}
  \addmilestone{at=phase-4.90,direction=90:2cm,text={IMT-2020 (5G) \\ \scriptsize Initial phase},text options={above, align=center}}
  \addmilestone{at=phase-5.90,direction=90:1cm,text={IMT-2020 (5G) \\ \scriptsize Second phase},text options={above, align=center}}
  \small
  \setlength{\baselineskip}{10pt}
  \addmilestone{at=phase-1.270,direction=270:2cm,text={Final \\channel models},text options={below, align=center,}}
  \addmilestone{at=phase-2.240,direction=250:1.2cm,text={Performance \\ evaluation \\ framework},text options={below, align=center,}}
  \addmilestone{at=phase-2.270,direction=270:1.8cm,text={Spectrum \\ scenarii},text options={below, align=center,}}
  \addmilestone{at=phase-2.310,direction=270:1cm,text={Performance \\ evaluation \\ results},text options={below, align=center,}}
  \addmilestone{at=phase-3.240,direction=240:1cm,text={Study on high bands \\ channel models},text options={below, align=center,}}
  \addmilestone{at=phase-3.270,direction=270:2.5cm,text={\textbf{Channel measurements} \\ (all partners)},text options={below, align=center,}}
  \addmilestone{at=phase-4.285,direction=270:3.5cm,text={Evaluation criteria \\ \& methods},text options={below, align=center,}}
  \addmilestone{at=phase-5.230,direction=270:3cm,text={New proposals},text options={below, align=center,}}
  \addmilestone{at=phase-5.270,direction=270:1cm,text={Evaluation},text options={below, align=center,}}
  \addmilestone{at=phase-5.310,direction=270:2.2cm,text={Outcome \\ \textbf{Start of the specification drafts}},text options={below, align=center,}}

\end{tikzpicture}
	\caption{Timeline of recent and upcoming milestones in channel modeling, with links to the standardization process, as planned in the last 3GPP Radio Access Network (RAN) group meetings. As of this publication, most partners in the 3GPP will have engaged in large measurement and validation campaigns related to the issues and challenges identified in this communication. As one can see, 2016 is pivotal in the number of measurement efforts, as well as the scenarios and evaluation framework definition for the future 5G standards.}
	\label{fig:timeline}
\end{figure}

Beyond the large amount of measurements to parametrize the models, the models' dominating factors are still unclear in new scenarios such as mmWave and high-speed ground transportation.
This requires further measurements campaigns before the actual model validations can start.
From the technological drivers described here, we identified the key environmental parameters that we expect will influence the performance of future networks in Sec.~\ref{sec:trends}.
This definition process is already taking place, as we can see in Fig.~\ref{fig:timeline}.
Large partnerships are aiming toward the 5G standardization process.
The year 2016 will see a number of channel measurement efforts in 3GPP, led by all the partners, as discussed in the last meeting of the RAN work group.
In parallel, both the new METIS-II project, as well as the IMT-2020 partnership will produce their standard evaluation scenarios, which will precisely define the needs and expectations of future RANs with respect to their simulation models. 

Finally, let us reconsider the primary of channel modeling for wireless telecommunications: providing a foundation for the design of disruptive technologies.
In parallel with improved simulation tools, there is still a need to capture these essential channel behaviors into simpler design models used by more theoretical researchers.
These models need to be ahead of the standardization curve.
In general, the remaining open problems similar to those expressed in this communication, with a small set of parameters and greater generality.
For example, as of today, there are no usable design models for D-MIMO that accurately reproduce the joint evolution of links between different access points and specific users.
The community also lacks more precise measurements and analysis of the stationarity intervals of channels in time, frequency and space -- a discussion that is related to the joint behavior of links over these dimensions.
As an accurate estimation of the channel statistics is a requirement for advanced communication algorithms, stationarity is an important operational factor.

As large measurement campaigns are performed to move toward 5G, this data may be used to extract new design models and operational bounds, which will be the basis of tomorrow's technological breakthroughs.

\bibliographystyle{IEEEtran}

% \clearpage

\end{document}